\documentclass[twocolumn,prd,aps,showpacs,amsmath,amssymb]{revtex4-1}
\usepackage{graphics}
\usepackage{graphicx}
\usepackage{dcolumn}
\usepackage{bm}
\usepackage{mathrsfs}
\usepackage{pstricks}
\usepackage{color}
\usepackage{slashed}
\usepackage{amsmath}
\usepackage{epsfig}
\usepackage{amsfonts}
\usepackage{amssymb}

\def\beq{\begin{equation}}
\def\eeq{\end{equation}}
\def\bea{\begin{eqnarray}}
\def\eea{\end{eqnarray}}
\def\nn{\nonumber}
\begin{document}

\title{Thermal and nonthermal scaling of the Casimir-Polder interaction\\ 
in a black hole spacetime}
\author{Gabriel Menezes}
\email{gabrielmenezes@ufrrj.br}
\affiliation{Grupo de F\'isica Te\'orica e Matem\'atica F\'isica,
  Departamento de F\'isica, Universidade Federal Rural do Rio de
  Janeiro, 23897-000 Serop\'edica, RJ, Brazil} 

\author{Claus Kiefer}

\email{kiefer@thp.uni-koeln.de}

\author{Jamir Marino}

\email{jmarino@thp.uni-koeln.de}

\affiliation{Institut f\"{u}r Theoretische Physik, Universit\"{a}t zu
  K\"{o}ln, Z\"ulpicher Stra\ss e 77, 50937 Cologne, Germany} 
%

\begin{abstract}
We study the Casimir-Polder force arising between two identical
two-level atoms and mediated by a massless scalar field propagating in
a black-hole background.  We study the interplay of  Hawking radiation
and  Casimir-Polder forces and find that, when the atoms are placed
near the event horizon, the scaling of the Casimir-Polder interaction
energy as a function of interatomic distance displays a transition
from a thermallike character  to a nonthermal behavior. We
corroborate our findings for a quantum field  prepared in the
Boulware, Hartle-Hawking, and Unruh vacua.  
Our analysis is consistent with the nonthermal character of the
Casimir-Polder interaction of two-level atoms in a relativistic
accelerated frame [J. Marino \emph{et al.},  Phys. Rev. Lett. 113,
020403 (2014)], where a crossover from thermal scaling, consistent
with the Unruh effect, to a nonthermal scaling has been observed. 
The two crossovers are a consequence  of the noninertial character of
the  background where the field mediating the Casimir interaction
propagates. 
While in the former case the characteristic crossover length scale is
proportional to the inverse of the surface gravity of the black hole,
in the latter it is determined by the inverse of the proper
acceleration of the atoms. 

\end{abstract}


\maketitle

\section{Introduction}
\label{intro}

The relationship between quantum theory and the gravitational field is
a very special one \cite{oup}. While standard quantum (field) theory is 
formulated on a fixed background, gravity is described by a
dynamical spacetime. This difference is the major obstacle for a
consistent quantization.  

Black holes are assumed to play a key role in the search for a quantum
theory of gravity. This is because they obey laws which are closely
analogous to the laws of thermodynamics. The interpretation of these
laws necessarily invokes quantum theory because classically a black
hole cannot radiate and thus cannot be attributed a temperature. Using
the formalism of quantum field theory on a classical dynamical
spacetime (see e.g. \cite{birrel,frolov}), Hawking has shown that black holes
radiate with a temperature proportional to $\hbar$, which in the case
of a Schwarzschild black hole is given by $T=\hbar c^3/8\pi k_{\rm
  B}GM$~\cite{hawking}. 
 
The interpretation and consequences of this temperature are still
subject of investigations; see, for example, \cite{Jerusalem} for a
recent review. If this radiation were exactly thermal and if the black
hole evaporated completely, any initial (quantum) state would evolve
into the same final thermal state, in violation of the unitary time
evolution of ordinary quantum theory. This ``information-loss
problem'' can eventually only be solved within a final theory of
quantum gravity. 

Besides particle creation, an important effect in black-hole
spacetimes is vacuum polarization \cite{frolov}. A similar effect in
flat spacetime occurs in the presence of nontrivial boundaries -- the
Casimir effect \cite{casimir,plunien,grib,bordag,book,Milonni}. The
physical reality of this effect has been empirically confirmed in a variety
of experiments \cite{L07}. 
In its classic formulation, the Casimir effect is the attraction of two
neutral conducting plates at zero temperature as a result
of a quantum pressure induced by vacuum fluctuations.  
Even before, Casimir and Polder have investigated the attraction
between an atom and a perfectly conducting wall as well as between two
atoms \cite{CP48}. 
Both Casimir and Casimir-Polder forces have been explored in the
presence of boundary conditions and nontrivial backgrounds which can
modify the quantization conditions of the field modes and accordingly
the structure of correlations in the quantum vacuum. See also Ref.~\cite{farina} 
for a recent study of the Casimir-Polder interaction in graphene.

In this paper, we address the Casimir-Polder interaction between two atoms. 
When embedded in a quantum vacuum, they 
experience a force as a result of local dipoles spontaneously induced
on them by correlated zero-point  vacuum fluctuations. We
consider this interaction in a black-hole spacetime and thus present
 a situation in which the quantum aspects of
black holes {\em and} the quantum aspects of the standard
Casimir-Polder force are intertwined. In this respect, we also remark the existence of 
a number of previous studies on the gravitation interaction of 
the Casimir energy~\cite{milton}.

A useful technique, widely employed in the literature, is a method
developed by Dalibard, Dupont-Roc, and Cohen-Tannoudji (DDC) in order
to separate in perturbation theory the distinct contributions of
vacuum fluctuations and radiation reaction to radiative shifts of
atomic energy levels~\cite{cohen2}.  
The method was originally formulated to treat a small system coupled to a reservoir \cite{cohen3}; in this case, it was shown that two types of physical processes contribute to the evolution of an observable, those
where the fluctuation of the reservoir polarizes the system  and those where it is the system itself that polarizes the reservoir. If the system is a quantized field, we call the former vacuum fluctuations and the latter radiation reaction contributions.

In second order perturbation theory, the aforementioned method has been
successfully applied to computing space-dependent radiative shifts of atoms in
front of a reflecting plate -- also known as atom-plate Casimir
force~\cite{Rizzuto}. 
Moreover, it has also been employed to investigate the radiative
processes of entangled atoms in Minkowski spacetime~\cite{ng1} and also in the presence of
an event horizon~\cite{ng2,ng3}. 

However, only recently the method by DDC has been extended to fourth order in perturbation theory for atoms
linearly coupled to a scalar field~\cite{marino, noto}, as necessary to
compute the Casimir-Polder force between two polarizable, neutral atoms in their respective
ground states.

 \subsection*{Outline of results}
 
\emph{(i) Casimir-Polder interaction in a black-hole spacetime --}
We examine the Casimir-Polder interaction between two identical
two-level atoms in a Schwarzschild spacetime, linearly coupled with
the quantum fluctuations of a scalar field prepared in the Boulware, Hartle-Hawking, and Unruh
vacuum states.  
Similar computations have already been performed with a different
atomic configuration, considering an electromagnetic field and
employing the method of equal-time spatial vacuum field
fluctuations~\cite{bh}. In this paper, we explore a broader variety of
parameters, considering the interplay of the interatomic distance,  energy level spacing
of the atoms, and the surface gravity of the black hole. 
In particular, we highlight the regimes where the Casimir-Polder force
exhibits a nonthermal scaling with the interatomic distance; indeed,
while for certain choices of parameters, the Casimir interaction
displays a thermal character linked to its Hawking temperature, at
large enough interatomic separations, and close to the black-hole
horizon, the noninertial character of the background metric modifies
the scaling of the force in a nonthermal fashion. 

We derive these results calculating the vacuum fluctuation and radiation reaction contributions to the Casimir-Polder interaction at fourth order in perturbation theory in the atom-field coupling strength (see for a derivation ~\cite{noto}).
The vacuum fluctuation term can be interpreted as the fluctuations of
the zero-point field inducing local dipoles on the atoms, which leads
to a coupling between the atoms,
while the radiation reaction term reflects the opposite
mechanism: when one of the atom experiences quantum fluctuations, it
polarizes the remainder 
of the system (the field and the other atom). 
The associated expressions for the radiative energy level shifts
(Eqs.~(3-4) below and Refs.~\cite{marino, noto}) provide a set of  general formulae to compute
Casimir-Polder forces from first principles without resorting to
specific phenomenological models.

\emph{(ii) Analogy with the Casimir-Polder interaction of two
  relativistic uniformly accelerated atoms --} We discuss the analogy
with a similar phenomenology encountered in Ref.~\cite{marino} (also
notice similar studies in~\cite{rizzutoUnruh}), where the large-distance
scaling of the Casimir interaction among two relativistic
uniformly accelerated atoms was studied. 
Close to the event horizon, the Schwarzschild metric takes the form of
the Rindler line element, and we find that the characteristic exponent
of the algebraic scaling discussed in~\cite{marino} is perfectly
mirrored in the large interatomic separation scaling of the Casimir
force close to the black hole. The nonthermal correction to the
Casimir interaction is imprinted by the noninertial character of the
background metric which becomes sizeable at distances larger than the
inverse of the surface gravity of the black hole, that is, larger than
$4GM/c^2$.

The organization of the paper is as follows. In Sec.~\ref{model}, we
setup our system and discuss the identification of vacuum
fluctuations and radiation reaction corrections at fourth order in
perturbation theory to the radiative energy shifts of the atoms. In
Sec.~\ref{calculation}, we calculate the Casimir-Polder interaction
energy for static atoms outside a Schwarzschild black hole and 
compare it with analogous results for relativistically uniformly
accelerated atoms \cite{marino}. Conclusions and final remarks are given in
Sec.~\ref{conclude}.  
In the Appendix, we present the correlation functions for a scalar field
in Schwarzschild spacetime. 

In this paper, we use units such that
$\hbar = c = k_B = 1$, but include some remarks on the dependence
of the results on such constants. We employ the convention that the Minkowski
signature is given by $\eta_{\alpha\beta} = -1, \alpha=\beta=1,2,3$,
$\eta_{\alpha\beta} = 1, \alpha=\beta=0$ and $\eta_{\alpha\beta} =
0,\alpha \neq \beta$.

\section{The model and the method}
\label{model}

In the following, we consider two identical two-level atoms
interacting with a quantum massless scalar field. The atoms move along
different world lines in a four-dimensional Schwarzschild spacetime, 
\begin{equation}
ds^2 = g_{00}dt^2 -{g^{-1}_{00}}dr^2 -r^2 (d\theta^2 + \sin^2\theta d\phi^2), 
\label{schw}
\end{equation}
where $g_{00}=\left(1 - r_{s}/r\right)$ and  $r_{s} = 2GM$ is the Schwarzschild radius.
Eq. \eqref{schw} describes the gravitational field outside a spherically symmetric body
of mass $M$ in spherical coordinates $(r,\theta,\phi)$. The collapse
of an electrically neutral, static star endowed with spherical
symmetry produces a spherical black hole of mass $M$ with external
gravitational field, described by the Schwarzschild line
element~(\ref{schw}), and with the event horizon of the black hole
being located at the Schwarzschild radius $r_{s}$.   

Our goal is to compute the Casimir-Polder force between the two
atoms mediated by a massless scalar field propagating in a spacetime
described by the metric
\eqref{schw}. We employ a general method for the computation of
Casimir-Polder interaction energy from first principles. The approach
we use is the DDC formalism up to fourth order in perturbation theory,
following closely  Refs.~\cite{marino, noto}. In
particular, we consider the contribution to the interaction energy
coming from the interplay between vacuum fluctuations and radiation
reaction among the two identical two-level atoms linearly coupled
with the  scalar field. We assume that both atoms are moving along
different stationary trajectories $x^{\mu}(\tau_i) = (t(\tau_i),{\bf
  x}(\tau_i))$, where $\tau_i$ denotes the proper time of atom
$i$ ($i=A, B$). The Hamiltonian of the system reads~\cite{Rizzuto}
\bea
H(t) &=& \omega_0\,\sigma_{3}^{A}\,\frac{d\tau_A}{dt} +
\omega_0\,\sigma_{3}^{B}\,\frac{d\tau_B}{dt}  
+ \sum_{{\bf k}}\, \omega_{{\bf k}}\,a^{\dagger}_{{\bf k}}(t)a_{{\bf k}}(t) 
\nn\\ 
&+& \lambda\sigma_{2}^{A}\,\varphi(x_{A}(\tau_{A}))\,\frac{d\tau_A}{dt} 
+ \lambda\sigma_{2}^{B}\,\varphi(x_{B}(\tau_{B}))\,\frac{d\tau_B}{dt},
\label{casimir-scalar} 
\eea
where $d\tau/dt  = \sqrt{g_{00}} $, and $t$ is the Schwarzschild
coordinate time. In Eq.~\eqref{casimir-scalar}, $\sigma_{3}^{j} =
1/2\left(|e_{j}\rangle\langle e_{j}| -   
|g_{j}\rangle\langle g_{j}|\right)$, $j = A, B$, $\omega_{{\bf k}} =
|{\bf k}|$, $a^{\dagger}_{{\bf k}}, a_{{\bf k}}$ are the usual
creation and annihilation operators of the scalar field quanta with
momentum $\bf k$.  The states $|g_A\rangle$, $|g_B\rangle$ and
$|e_A\rangle$, $|e_B\rangle$ denote the ground and excited states of
isolated atoms with energies $-\omega_0/2$ and $\omega_0/2$,
respectively. One can write $\sigma_{2}^{j} = (i/2)(\sigma_{-}^{j} -
\sigma_{+}^{j})$, $j = A, B$, where $\sigma_{\pm}^{j}$ are the usual
atomic raising and lowering operators, satisfying the algebra:
$[\sigma_{3}^{i},\sigma_{\pm}^{j}] = \pm \sigma_{\pm}^{i}\delta^{ij}$
and  
$[\sigma_{+}^{i},\sigma_{-}^{j}] = 2\sigma_{3}^{i}\delta^{ij}$.
Finally, $\lambda$ is the light-matter coupling strength. 

A brief comment on the units employed in this work is in order. Here the field $\varphi$ has the usual mass
dimension, while $\lambda$ is dimensionless. If we reinsert $\hbar$, $\varphi^2$
 has dimension of mass over length, while $\lambda^2$ has dimension of
 mass times length.

The DDC approach allows us to identify two different contributions in the
expectation value of a given atomic observable~\cite{cohen2, cohen3}; the first is generally
refered to as \emph{vacuum fluctuation} (vf) term and it accounts for
the response of the atom to zero-point quantum fluctuations of the
field, while the other term accounts for the backreaction on the atom,
as a result of its interaction with the field -- it is the \emph{radiation
  reaction} (rr).  
Here, we do not give a detailed treatment of the DDC formalism, since
it has been  discussed to full extent in the papers cited 
above; we will directly present, instead, the final outcome of
the derivation of a fourth-order perturbative computation  ($\lambda$
is the small parameter) of the \emph{vf} and \emph{rr} contributions
to the radiative correction of the atomic bare energy, $\omega_0$, of
a given atom~\cite{marino, noto}. In particular, in order to extract  the Casimir-Polder
interaction, the part of energy shift of interest is the contribution
at fourth order in the atom-field interaction that depends on the
interatomic distance (since the other fourth-order terms are just
renormalizations of the bare energy $\omega_0$).  

We focus, for instance, on the radiative shift to the level
$|\alpha\rangle$ of atom $A$, and  we take the average of field
operators in the vacuum state of the quantum field $|0\rangle$ as well
as the expectation value of atomic operators in the  state
$|\nu\rangle$ of  atom $B$. After a lengthy algebra, we find the
following expression for the vacuum-fluctuation and radiation-reaction
contributions to the energy level shift of atom $A$ in the state
$|\alpha\rangle$:  
\begin{widetext}
\bea
(\delta E)^{A}_{\alpha,\textrm{vf}} &=& \frac{i \lambda^4}{4}\,\int_{t_0}^{t}dt'\,\int_{t_0}^{t'}dt''\,\int_{t_0}^{t''}dt'''\,
D[x_{A}(\tau_{A}(t)),x_{B}(\tau_{B}'''(t''')))]\Delta[x_{A}(\tau_{A}'(t')),x_{B}(\tau_{B}''(t'')))]\,
\nn\\
&\times&\,\chi^{A}_{\alpha}[\tau_A(t),\tau_A'(t')]\,\chi^{B}_{\beta}[\tau_B''(t''),\tau_B'''(t''')]
\frac{d\tau_A}{dt}\frac{d\tau_A'}{dt'}\frac{d\tau_B''}{dt''}\frac{d\tau_B'''}{dt'''}
\label{vf-shift}
\eea
and
\bea
(\delta E)^{A}_{\alpha,\textrm{rr}} &=& -\frac{i \lambda^4}{4}\,\int_{t_0}^{t}dt'\,\int_{t_0}^{t'}dt''\,\int_{t_0}^{t''}dt'''\,
\Delta[x_{A}(\tau_{A}(t)),x_{B}(\tau_{B}'(t')))]\Delta[x_{A}(\tau_{A}'''(t''')),x_{B}(\tau_{B}''(t'')))]\,
\nn\\
&\times&\,C^{A}_{\alpha}[\tau_A(t),\tau_A'''(t''')]\,\chi^{B}_{\beta}[\tau_B'(t'),\tau_B''(t'')]
\frac{d\tau_A}{dt}\frac{d\tau_B'}{dt'}\frac{d\tau_B''}{dt''}\frac{d\tau_A'''}{dt'''}.
\label{rr-shift}
\eea
\end{widetext}

In all equations above, we make use of the following definitions:
\beq
\chi^{kl}_{\nu}(t,t') := \langle \nu | [\sigma_{2}^{k,f}(\tau_{k}(t)),\sigma_{2}^{l,f}(\tau_{l}(t'))]| \nu \rangle,
\label{susa}
\eeq
($\chi^{kk}_{\nu} \equiv \chi^{k}_{\nu}$) $k,l = A, B$, is the atomic susceptibility of the atom in the state $|\nu\rangle$, and
\beq
C^{kl}_{\nu}(t,t') := \langle \nu | [\sigma_{2}^{k,f}(\tau_{k}(t)),\sigma_{2}^{l,f}(\tau_{l}(t'))]| \nu \rangle,
\eeq
($C^{kk}_{\nu} \equiv C^{k}_{\nu}$) is the symmetric correlation
function of the atom in the state $|\nu\rangle$. (The suffix $f$
indicates that we are in the interacton picture where we have the
{\em free} evolution of the atomic observables.)

The explicit forms of these quantities are given by 
\bea
\chi^{ab}_{\nu}(t,t') &=& \sum_{\nu'} \biggl[{\cal A}^{ab}(\nu, \nu')\,e^{i\Delta\nu(\tau_a(t) - \tau_b'(t'))} 
\nn\\
&-&\,{\cal A}^{ba}(\nu, \nu')\,e^{-i\Delta\nu(\tau_a(t) - \tau_b'(t'))}\biggr],
\label{susa1-scalar}
\eea
and
\bea
C^{ab}_{\nu}(t,t') &=& \sum_{\nu'} \biggl[{\cal A}^{ab}(\nu, \nu')\,e^{i\Delta\nu(\tau_a(t) - \tau_b'(t'))} 
\nn\\
&+&\, {\cal A}^{ba}(\nu, \nu')\,e^{-i\Delta\nu(\tau_a(t) - \tau_b'(t'))}\biggr],
\label{cora1-scalar}
\eea
where $\Delta \nu :=\omega_0( \nu - \nu')$ (the summation over $\nu'$
is over the product state basis or, in the case of one atom, over
$|g\rangle,|e\rangle$), and we have conveniently introduced the
function ${\cal A}_{ab}^{ij}(\nu, \nu')$ defined as 
\beq
{\cal A}^{ab}(\nu, \nu') =  \langle \nu |\sigma_{2}^{a,f}(0)| \nu' \rangle\langle \nu' |\sigma_{2}^{b,f}(0)| \nu \rangle.
\label{aa}
\eeq
For the field variables, one has
\beq
D(x(\tau),x(\tau')) = \langle 0 |\{\varphi^{f}(x(\tau)),\varphi^{f}(x(\tau'))\}| 0 \rangle,
\label{hada}
\eeq
which is the symmetric correlation function of the scalar field (also
known as Hadamard's elementary function) and  
\beq
\Delta(x(\tau),x(\tau')) = \langle 0 |[\varphi^{f}(x(\tau)),\varphi^{f}(x(\tau'))]| 0 \rangle,
\label{pauli}
\eeq
which is the response function of the field (or Pauli-Jordan function). 

The physical interpretation of Eqs.~\eqref{vf-shift}
and~\eqref{rr-shift} can be read from the type of response or
correlation functions entering these expressions. Regarding $\delta
E_{vf}$, the field fluctuates around the two atoms $A$ and $B$
($D_{AB}$), and they respond with a local polarization $\chi^{A}$ and
$\chi^B$, which results in the transmission of a quantum of the field
between them (response field, $\Delta_{AB}$), or in other words, the
medium among them gets polarized. Regarding $\delta E_{rr}$, the atom
$A$ fluctuates ($C^A$), and polarizes the remaining components of the
system: the atom $B$ ($\chi^B$) and the field ($\Delta_{AB}$). 

\section{Casimir-Polder interaction}
\label{calculation}

We consider the two atoms prepared in their respective ground states,
static and at fixed Schwarzschild radial coordinates $r_{A}$ and
$r_{B}$ outside the  black hole. The world lines are given
respectively by $x^{\mu}(\tau_i) = (\tau_i/\sqrt{g_{00(r_i)}} , r_i,
\theta_i, \phi_i)$, $i = A, B$, and $g_{00}(r) = 1 - 2GM/r$ (the angular
coordinates $\theta_i, \phi_i$ are constants). In order to employ the
formulae~\eqref{vf-shift} and \eqref{rr-shift}, one should use the
associated 
correlation functions of the scalar field. The correlation functions
of a massless scalar field in Schwarzschild spacetime for each one of
the possible vacua (Boulware, Hartle-Hawking, Unruh) discussed in the
literature is briefly outlined in the Appendix (where we present the
computation of the correlations functions relevant to extract $\Delta$
and $D$). For further details, we refer to Ref.~\cite{frolov}.

\subsection{Boulware vacuum}
\label{Boulware}

The Boulware vacuum has a close similarity to the  concept of an empty
state at large radii. It is the appropriate choice of vacuum state for
quantum fields in the vicinity of an isolated, cold neutron star;  the
Boulware vacuum is relevant to the exterior region of a massive body
that is  just outside its Schwarzschild radius~\cite{boul,sciama}. 

The associated symmetric correlation function is given by
\begin{widetext}
\bea
D^{B}(x_i(t),x_j(t')) &=& \frac{1}{16\pi^2}\sum_{l = 0}^{\infty}(2l
+1)P_{l}(\hat{r}_i \cdot \hat{r}_j) 
\int_{0}^{\infty}\,\frac{d\omega}{\omega}
\,\Biggl\{e^{-i\omega (t-t')}\left[\overrightarrow{R}_{\omega
    l}(r_i)\overrightarrow{R}^{*}_{\omega l}(r_j) +
  \overleftarrow{R}_{\omega l}(r_i)\overleftarrow{R}^{*}_{\omega
    l}(r_j)\right] 
\nn\\
&+&\, e^{i\omega (t-t')}\left[\overrightarrow{R}_{\omega
    l}(r_j)\overrightarrow{R}^{*}_{\omega l}(r_i) +
  \overleftarrow{R}_{\omega l}(r_j)\overleftarrow{R}^{*}_{\omega
    l}(r_i)\right]\Biggr\}, 
\label{hada-rest-bouls}
\eea
where the addition theorem for the spherical harmonics was used~\cite{hilbert}, $\hat{r}_{i}$ and $\hat{r}_{j}$ (with $i, j =A, B$) are two unit vectors with spherical coordinates $(\theta_{i}, \phi_{i})$ and $(\theta_{j}, \phi_{j})$, respectively,  $P_{l}$ is the Legendre polynomial of degree $l$~\cite{abram} and the radial functions $\overrightarrow{R}$ and $\overleftarrow{R}$ are introduced in the Appendix. The response function is given by
\bea
\Delta^{B}(x_i(t),x_j(t')) &=& \frac{1}{16\pi^2}\sum_{l = 0}^{\infty}(2l +1)P_{l}(\hat{r}_i \cdot \hat{r}_j)
\int_{0}^{\infty}\,\frac{d\omega}{\omega}
\,\Biggl\{e^{-i\omega (t-t')}\left[\overrightarrow{R}_{\omega l}(r_i)\overrightarrow{R}^{*}_{\omega l}(r_j) + \overleftarrow{R}_{\omega l}(r_i)\overleftarrow{R}^{*}_{\omega l}(r_j)\right]
\nn\\
&-& e^{i\omega (t-t')}\left[\overrightarrow{R}_{\omega l}(r_j)\overrightarrow{R}^{*}_{\omega l}(r_i) + \overleftarrow{R}_{\omega l}(r_j)\overleftarrow{R}^{*}_{\omega l}(r_i)\right]\Biggr\}.
\label{pauli-rest-bouls}
\eea
In this way, using ${\cal A}^{AA}(\nu, \nu') = {\cal A}^{BB}(\nu,
\nu') = 1/4$, with $|\nu\rangle=|g\rangle$  being the atomic ground state
$|g\rangle$ (the only contribution in the summation over $\nu'$
appearing in the atomic correlation functions comes from the excited
state $|\nu'\rangle = |e\rangle$), one has, for the vacuum fluctuation
contribution
\bea
(\delta E)^{A}_{g,\textrm{vf}} &=& \frac{i \lambda^4}{2^{14}\pi^4}\,g_{00}(r_{A})g_{00}(r_{B})
\int_{0}^{\infty}\,\frac{d\omega}{\omega}\int_{0}^{\infty}\,\frac{d\omega'}{\omega'}
\int_{t_0}^{t}dt'\,\int_{t_0}^{t'}dt''\,\int_{t_0}^{t''}dt'''
\nn\\
&\times&\,\left[e^{-i\omega (t-t''')}B(\omega, r_A, r_B) + e^{i\omega (t-t''')}B(\omega, r_B, r_A)\right]
\nn\\
&\times&\,\left[e^{-i\omega' (t'-t'')}B(\omega', r_A, r_B) - e^{i\omega' (t'-t'')}B(\omega', r_B, r_A)\right]
\nn\\
&\times&\, \biggl[e^{i\omega_{A}(t - t')} - e^{-i\omega_{A}(t - t')}\biggr]
\biggl[e^{i\omega_{B}(t'' - t''')} - e^{-i\omega_{B}(t'' - t''')}\biggr],
\eea
where $\omega_A = -\sqrt{g_{00}(r_A)}\,\omega_0$, $\omega_B = -\sqrt{g_{00}(r_B)}\,\omega_0$. The appearance of $\sqrt{g_{00}}$ multiplying the energy gap is a consequence of the usual gravitational redshift effect. In addition, we have defined
$B(\omega, r, r') = \overrightarrow{B}(\omega, r, r') + \overleftarrow{B}(\omega, r, r')$, with 
\bea
\overrightarrow{B}(\omega, r, r') &=& \sum_{l = 0}^{\infty} (2l+1)P_{l}(\hat{r} \cdot \hat{r}')
\overrightarrow{R}_{\omega l}(r)\overrightarrow{R}^{*}_{\omega l}(r'),
\nn\\
\overleftarrow{B}(\omega, r, r') &=& \sum_{l = 0}^{\infty} (2l+1)P_{l}(\hat{r} \cdot \hat{r}')
\overleftarrow{R}_{\omega l}(r)\overleftarrow{R}^{*}_{\omega l}(r').
\eea
Regarding the radiation reaction contribution, one gets
\bea
(\delta E)^{A}_{g,\textrm{rr}} &=& -\frac{i \lambda^4}{2^{14}\pi^4}\,g_{00}(r_{A})g_{00}(r_{B})
\,\int_{0}^{\infty}\,\frac{d\omega}{\omega}\int_{0}^{\infty}\,\frac{d\omega'}{\omega'}
\int_{t_0}^{t}dt'\,\int_{t_0}^{t'}dt''\,\int_{t_0}^{t''}dt'''
\nn\\
&\times&\,\left[e^{-i\omega (t-t')}B(\omega, r_A, r_B) - e^{i\omega (t-t')}B(\omega, r_B, r_A)\right]
\nn\\
&\times&\,\left[e^{-i\omega' (t'''-t'')}B(\omega', r_A, r_B) - e^{i\omega' (t'''-t'')}B(\omega', r_B, r_A)\right]
\nn\\
&\times&\, \biggl[e^{i\omega_{A}(t - t''')} + e^{-i\omega_{A}(t - t''')}\biggr]
\biggl[e^{i\omega_{B}(t' - t'')} - e^{-i\omega_{B}(t' - t'')}\biggr].
\eea

\subsubsection{Atoms far from the black hole}
\label{sub:far}
In order to keep the discussion transparent, let us discuss the
radiative energy shifts for the asymptotic regions of interest,
keeping $|{\bf x}_{A} - {\bf x}_{B}|$ fixed. First, we consider the
case $r_{A}, r_{B} \to \infty$. Following the discussion presented in
the Appendix, one can neglect the contribution coming from
$\overrightarrow{B}$. For $\overleftarrow{B}$, we get 
\bea
\overleftarrow{B}(\omega, r, r') &=& \sum_{l=0}^{\infty}(2l +
1)\,P_{l}(\hat{r} \cdot \hat{r}')\overleftarrow{R}_{\omega
  l}(r)\overleftarrow{R}^{*}_{\omega l}(r')  
\approx\, 4\omega\,\frac{\sin\left(\omega|{\bf x} - {\bf
      x}'|\right)}{|{\bf x} - {\bf x}'|},\,\,\, r, r' \to \infty. 
\eea
Hence:
\bea
(\delta E)^{A}_{g,\textrm{vf}} &\approx& -\frac{i\lambda^4}{16}\,\int_{t_0}^{t}dt'\,\int_{t_0}^{t'}dt''\,\int_{t_0}^{t''}dt'''
\,D^{M}(t-t''',{\bf x}_{A}-{\bf x}_{B})\Delta^{M}(t'-t'',{\bf x}_{A}-{\bf x}_{B})\,
\nn\\
&\times&\,\sin[\omega_{0}(t - t')]\,\sin[\omega_{0}(t'' - t''')],
\label{vf-infinity}
\eea
and
\bea
(\delta E)^{A}_{g,\textrm{rr}} &\approx& -\frac{\lambda^4}{16}\,\int_{t_0}^{t}dt'\,\int_{t_0}^{t'}dt''\,\int_{t_0}^{t''}dt'''
\,\Delta^{M}(t-t',{\bf x}_{A}-{\bf x}_{B})\Delta^{M}(t'''-t'',{\bf x}_{A}-{\bf x}_{B})\,
\nn\\
&\times&\,\cos[\omega_{0}(t - t''')]\,\sin[\omega_{0}(t' - t'')],
\label{rr-infinity}
\eea
where we have used the fact that $\omega_{A} \approx \omega_{B}
\approx -\,\omega_0$ for $r_{A}, r_{B} \to \infty$. In the above
expressions, we have used the definitions 
\bea
D^{M}(t-t',{\bf x}-{\bf x}') &=& \frac{1}{4\pi^2}\int_{0}^{\infty} d\omega\,\frac{\sin\left(\omega|{\bf x} - {\bf x}'|\right)}
{|{\bf x} - {\bf x}'|}\,\left(e^{-i\omega(t-t')} + e^{i\omega(t-t')}\right)
\nn\\
&=& -\frac{1}{4\pi^2}\left\{\frac{1}{(t-t'-i\epsilon)^2 -|{\bf x} - {\bf x}'|^2} 
+ \frac{1}{(t-t'+i\epsilon)^2 -|{\bf x} - {\bf x}'|^2}\right\},
\eea
and
\bea\label{eq:reg}
\Delta^{M}(t-t',{\bf x}-{\bf x}') &=& \frac{1}{4\pi^2}\int_{0}^{\infty} d\omega\,\frac{\sin\left(\omega|{\bf x} - {\bf x}'|\right)}
{|{\bf x} - {\bf x}'|}\,\left(e^{-i\omega(t-t')} - e^{i\omega(t-t')}\right)
\nn\\
&=& \frac{i}{4\pi|{\bf x} - {\bf x}'|}\left\{\delta[(t-t')+|{\bf x} - {\bf x}'|] - \delta[(t-t')-|{\bf x} - {\bf x}'|]\right\},
\eea
\end{widetext}
where $D^{M}$ and $\Delta^{M}$ have now the same expressions,
respectively, of  the symmetric correlation and the response functions
of the massless scalar field in Minkowski spacetime~\cite{marino}. Therefore, in the
limit $r_{A}, r_{B} \to \infty$, we recover the results for the scalar
Casimir-Polder energy between two static atoms in Minkowski
spacetime. Indeed, concerning vacuum fluctuations, one gets, in the
limit  
$t - t_0 \to \infty$,
\bea
(\delta E)^{A}_{g,\textrm{vf}} &\approx& -\frac{\lambda^4}{1024\pi^3\,\omega_0 (\Delta r)^2}
\nn\\
&\times&\,\Bigl[\pi\cos (2 \omega_0 \Delta r) + f(2\omega_0 \Delta r)\Bigr],
\eea
where $\Delta r = |{\bf x}_{A} - {\bf x}_{B}|$, and
\bea
f(z) &=& \pi\Bigl[-1 + z \sin (z)\Bigr] 
\nn\\
&+&\,2\text{Ci}(z)\Bigl[\sin (z) - z \cos (z)\Bigr]
\nn\\
&-&\,2\text{Si}(z)\Bigl[\cos(z) + z \sin (z)\Bigr];
\eea
$\text{Ci}$ and $\text{Si}$ are the usual cosine and sine integrals,
respectively. As for the radiation-reaction contribution, we find, in
the limit $t - t_0 \to \infty$, 
\beq
(\delta E)^{A}_{g,\textrm{rr}} \approx -\frac{\lambda^4\,\cos(2\omega _0 \Delta r)}{1024\pi^2\,\omega _0 (\Delta r)^2}.
\label{casimir-rr}
\eeq
In order to derive this expression we have considered a convergence factor
$e^{-\eta t''}$ (where $\eta$ is a positive infinitesimal) in the
integral over $t''$. This is required, since in the limit $\eta \to 0$
the integral over $t''$ diverges, as  expected for a nonrelativistic
evaluation of radiative energy shifts. This occurs also in the
calculation of Lamb shifts for static atoms within the DDC formalism
\cite{cohen2, cohen3}, and we have followed an analogous
regularization procedure here. In the final formulae of our
computations, we accordingly present only the finite part of
integrals. 

The total Casimir-Polder interaction energy is the sum of the above
contributions, 
\beq
\label{ECP1}
E_{\rm CP} := (\delta E)^{A}_{g,\textrm{vf}} + (\delta E)^{A}_{g,\textrm{rr}}.
\eeq
In the near-zone regime, $\omega_0 \Delta r \ll 1$, the leading order
is then given by the radiation-reaction contribution, specifically
\bea
\label{ECP2}
E_{\rm CP}\bigg|_{\omega_0 \Delta r \ll 1} &\approx& 
(\delta E)^{A}_{g,\textrm{rr}}\bigg|_{\omega_0 \Delta r \ll 1} 
\nn\\
&\approx&\, -\frac{\lambda^4}{1024\pi^2\,\omega _0 (\Delta r)^2}.
\eea

In the far-zone regime, $\omega_0 \Delta r \gg 1$, the leading behavior is due exclusively to the vacuum-fluctuation contribution; hence
\bea
E_{\rm CP}\bigg|_{\omega_0 \Delta r \gg 1} &\approx&
(\delta E)^{A}_{g,\textrm{vf}}\bigg|_{\omega_0 \Delta r \gg 1} 
\nn\\
&\approx& -\frac{\lambda^4}{512\pi^3\,\omega_0^2 (\Delta r)^3}.
\eea

As a benchmark, notice that these $1/(\Delta r)^2$ and $1/(\Delta r)^3$
scalings of the Casimir-Polder forces in the near ($\Delta
r\ll1/\omega_0$) and far zones ($\Delta r\gg1/\omega_0$),
respectively, were found
in Ref.~\cite{marino} for two static atoms in Minkowski
spacetime.  

If we want to reinsert the reduced Planck constant $\hbar$ back in these expressions, we have to
notice that $\lambda$ is then dimensionful, having the dimension of a
square root of mass times length. To compare it with the standard
expressions for the Casimir-Polder force \cite{CP48}, one has to
redefine the coupling as $\lambda^2=\hbar\tilde{\lambda}^2$, with
$\tilde{\lambda}$ being dimensionless. In this way, one recovers the
standard factor $\hbar$ (or $\hbar c$ if $c$ is taken into account) in
the numerator of these expressions.

\begin{widetext}

\subsubsection{Atoms close to the black hole}

We now  consider the limit $r_{A}, r_{B} \to 2GM$. In this situation, from the results derived in the Appendix, we have
\bea\label{eq:over}
\overrightarrow{B}(\omega, r, r') &=& \sum_{l = 0}^{\infty} (2l+1)P_{l}(\hat{r} \cdot \hat{r}')
\overrightarrow{R}_{\omega l}(r)\overrightarrow{R}^{*}_{\omega l}(r')
\nn\\
&\approx&\, \frac{4\omega }{\sqrt{g_{00}}}\,\frac{\sin\left[(2\omega/\kappa) \sinh^{-1}\left(\frac{\kappa(r) \Delta L}{2}\right)\right]}{\Delta L\sqrt{1 + (\kappa(r) \Delta L/2)^2}},
\,\,\,r, r' \to 2GM,
\eea
where $\kappa = 1/4GM$ is the surface gravity, $\kappa(r) =
\kappa/\sqrt{g_{00}(r)}$, and $\Delta L = r_s\gamma$ ($\cos\gamma = \hat{r}_{A} \cdot
\hat{r}_{B}$) is the arc distance between the atoms. In addition: 
\bea
\overleftarrow{B}(\omega, r, r') &=& \sum_{l=0}^{\infty}(2l + 1)\,P_{l}(\hat{r} \cdot \hat{r}')\overleftarrow{R}_{\omega l}(r)\overleftarrow{R}^{*}_{\omega l}(r') 
\nn\\
&\approx& \sum_{l=0}^{\infty}\frac{(2l+1)P_{l}(\hat{r} \cdot \hat{r}')
|{\cal T}_{l}(\omega)|^2 }{(2M)^2},\,\,\,r, r' \to 2GM.
\eea
This last expression can be neglected in comparison with Eq.~\eqref{eq:over} at  leading order in $r,r'\to2GM$. Therefore, for the \emph{vf} contribution to the energy level shift of the atom $A$, we find
\bea
(\delta E)^{A}_{g,\textrm{vf}} &\approx& -\frac{i\lambda^4}{16}\,g_{00}^2(r)
\int_{t_0}^{t}dt'\,\int_{t_0}^{t'}dt''\,\int_{t_0}^{t''}dt'''\,
\tilde{D}^{B}(t-t''',d_1,d_2)\,\tilde{\Delta}^{B}(t'-t'',d_1,d_2)
\nn\\
&\times&\,\sin[|\omega_{A}|(t - t')]\,\sin[|\omega_{A}|(t'' - t''')],
\label{vf-2m}
\eea
and, for the \emph{rr} term,
\bea
(\delta E)^{A}_{g,\textrm{rr}} &\approx& - \frac{\lambda^4}{16}\,g_{00}^2(r)
\int_{t_0}^{t}dt'\,\int_{t_0}^{t'}dt''\,\int_{t_0}^{t''}dt'''\,
\tilde{\Delta}^{B}(t-t',d_1,d_2)\,\tilde{\Delta}^{B}(t'''-t'',d_1,d_2)
\nn\\
&\times&\,\cos[|\omega_{A}|(t - t''')]\,\sin[|\omega_{A}|(t' - t'')],
\label{rr-2m}
\eea
where we have defined
\bea
\tilde{D}^{B}(t-t',{\bf x}-{\bf x}') &=& \frac{(g_{00})^{-1/2}}{4\pi^2\Delta L\sqrt{1 + (\kappa(r) \Delta L/2)^2}}
\,\int_{0}^{\infty} d\omega\,\sin\left(\omega d_{1}\right)\,\left(e^{-i\omega(t-t')} + e^{i\omega(t-t')}\right)
\nn\\
&=& -\frac{d_1\,(g_{00})^{-1/2}}{4\pi^2\Delta L\sqrt{1 + (\kappa(r) \Delta L/2)^2}}\left\{\frac{1}{(t-t'-i\epsilon)^2 - d_{1}^2} 
+ \frac{1}{(t-t'+i\epsilon)^2 - d_{1}^2}\right\},
\label{boul-hada-2m}
\eea
and
\bea
\tilde{\Delta}^{B}(t-t',{\bf x}-{\bf x}') &=& \frac{(g_{00})^{-1/2}}
{4\pi^2\Delta L\sqrt{1 + (\kappa(r) \Delta L/2)^2}}\int_{0}^{\infty} d\omega\,
\sin\left(\omega d_{1}\right)\,\left(e^{-i\omega(t-t')} - e^{i\omega(t-t')}\right)
\nn\\
&=&\, \frac{i\,(g_{00})^{-1/2}}{4\pi\Delta L\sqrt{1 + (\kappa(r) \Delta L/2)^2}}
\left\{\delta[(t-t')+ d_{1}] - \delta[(t-t')- d_{1}]\right\},
\label{boul-pauli-2m}
\eea
\end{widetext}
with
\beq
d_1(\gamma, r) = d_1 = \frac{2}{\kappa} \sinh^{-1}\left(\frac{\kappa(r) \Delta L}{2}\right).
\eeq
Above we used the fact that $\omega_{A} \approx \omega_{B}$  and $r_{A} \approx r_{B} = r$ for $r_{A}, r_{B} \to 2GM$ 
(but ${\bf \hat{r}}_A \neq {\bf \hat{r}}_B$). Therefore, proceeding with a similar calculation as the previous case one obtains the following expression for the contributions coming from the vacuum fluctuations, taking $t - t_0 \to \infty$ in Eqs.~\eqref{vf-2m} and~\eqref{rr-2m},
\bea
(\delta E)^{A}_{g,\textrm{vf}} &\approx& -\frac{\lambda^4}{1024\pi^3\,|\omega_A|}\,
\frac{g_{00}}{(\Delta L)^2\,(1 + (\kappa(r) \Delta L/2)^2)}
\nn\\
&\times&\,\Bigl[\pi\cos (2 |\omega_A| d_1)  + f(2 |\omega_A| d_1)\Bigr].
\eea
The (finite part of the) radiation-reaction contribution reads, in the limit $t - t_0 \to \infty$,
\bea
(\delta E)^{A}_{g,\textrm{rr}} &\approx& - \frac{\lambda^4\,g_{00}\,\cos(2\omega _A d_1)}
{1024\pi^2\,|\omega_A|\,(\Delta L)^2(1 + (\kappa(r) \Delta L/2)^2)}.
\label{casimir-rr-2m}
\eea
For $\kappa(r) \Delta L \ll 1$ (or $\gamma \ll 2\,\sqrt{g_{00}}$), we have
$d_1 \approx \Delta L/\sqrt{g_{00}}$ and  
 $(\Delta L)^2\,(1 + (\kappa(r) \Delta L/2)^2) \approx (\Delta L)^2$;
 therefore, we obtain similar results as for the case of atoms placed at
 $r_A,r_B\to\infty$, taking into account the necessary changes coming from gravitational-redshift effects. \\ 
 
On the other hand, for the more realistic situation in which
$\kappa(r) \Delta L \gg 1$ (or $\gamma \gg 2\,\sqrt{g_{00}}$), since
$g_{00}$ is a small quantity near the event horizon, one finds that
$d_1 \approx 2 \ln[\kappa(r) \Delta L]/\kappa$. Assuming the energy
spacing of the atoms to be larger than the surface gravity $|\omega_{A}|/\kappa \gg 1$, one has, for the vacuum-fluctuation contribution at  leading order,
\bea
(\delta E)^{A}_{g,\textrm{vf}}\bigg|_{\kappa(r) \Delta L \gg 1} &\approx& -\frac{\lambda^4 \kappa^3}{256\pi^3\,
|\omega_A|^2}\,\frac{1}{(\kappa(r) \Delta L)^4\,\ln(\kappa(r) \Delta L)}
\nn\\
&\times&\,\left(1 -\frac{\pi|\omega_A|}{\kappa} \ln(\kappa(r) \Delta L)\right).
\eea
In order to derive this result, one first has to develop an asymptotic series
in $|\omega_{A}|/\kappa$  and then expand  in $\kappa(r) \Delta
L\gg1$. In a similar fashion, the finite part of the contribution
coming from radiation reaction reads 
\bea
(\delta E)^{A}_{g,\textrm{rr}}\bigg|_{\kappa(r) \Delta L \gg 1} &\approx& -\frac{\lambda^4\,\kappa^2}{256\pi^2\,|\omega_A|}\,
\frac{1}{(\kappa(r) \Delta L)^4}.
\label{casimir-rr-2m2}
\eea
Therefore, in the limit $r \to 2GM$ and keeping  $\kappa(r) \Delta L \gg 1$ and $|\omega_{A}|/\kappa \gg 1$, the Casimir-Polder energy reads%
\bea
E_{\rm CP} &=& (\delta E)^{A}_{g,\textrm{vf}} + (\delta E)^{A}_{g,\textrm{rr}} 
\nn\\
&\approx& -\frac{\lambda^4 \kappa^3}{256\pi^3\,
|\omega_A|^2}\,\frac{1}{(\kappa(r) \Delta L)^4\,\ln(\kappa(r) \Delta L)}. 
\eea
We emphasize that this result differs strongly from
the setup with two atoms far away from the black hole as discussed in subsection
\ref{sub:far} above; the power-law scaling of the Casimir-Polder
interaction energy is clearly different. We also note that
the gravitational constant explicitly occurs in
these expressions (through $\kappa$), in contrast to the ealier
results \eqref{ECP1} and \eqref{ECP2}.  

The different scaling of the Casimir interaction energy at large
distances $\kappa(r)\Delta L\gg1$ is due to corrections proportional
to $\Delta L$ in the two-point response and correlation functions, and
it signals the fact that at large enough distances the strong
noninertial character of the metric becomes pronounced; on the
contrary, at short distances, they are negligible and the Casimir
interaction is then well approximated by its expression in flat
spacetime (Eq.~\eqref{casimir-rr-2m} and discussion below). We believe
that this characteristic scaling of the Casimir energy can have
important consequences in the situation in which matter is around a body collapsing
towards its Schwarzschild radius during the evolution towards  a black
hole. 

\subsection{Hartle-Hawking vacuum}
\label{Hartle-Hawking}

The Hartle-Hawking vacuum is relevant for the physical situation in which the black hole is at equilibrium with black-body radiation at temperature $T=\kappa/(2\pi)$~\cite{sciama,haw}.

The associated symmetric correlation function is given by
\begin{widetext}
\bea
D^{H}(x_i(t),x_j(t')) &=& \frac{1}{16\pi^2}\sum_{l = 0}^{\infty}(2l +1)P_{l}(\hat{r}_i \cdot \hat{r}_j)
\int_{-\infty}^{\infty}\,\frac{d\omega}{\omega}
\,\Biggl\{e^{-i\omega (t-t')} 
\left[\frac{\overrightarrow{R}_{\omega l}(r_i)\overrightarrow{R}^{*}_{\omega l}(r_j)}{1 - e^{-2\pi\omega/\kappa}} + \frac{\overleftarrow{R}_{\omega l}(r_i)\overleftarrow{R}^{*}_{\omega l}(r_j)}{e^{2\pi\omega/\kappa} - 1}\right]
\nn\\
&& + \,e^{i\omega (t-t')} 
\left[\frac{\overrightarrow{R}_{\omega l}(r_j)\overrightarrow{R}^{*}_{\omega l}(r_i)}{1 - e^{-2\pi\omega/\kappa}} + \frac{\overleftarrow{R}_{\omega l}(r_j)\overleftarrow{R}^{*}_{\omega l}(r_i)}{e^{2\pi\omega/\kappa} - 1}\right]\Biggr\},
\label{hada-rest-haws}
\eea
where again we have employed the addition theorem for  spherical harmonics,  while for the response function, we have 
\bea
\Delta^{H}(x_1i(t),x_j(t')) &=& \frac{1}{16\pi^2}\sum_{l = 0}^{\infty}(2l +1)P_{l}(\hat{r}_i \cdot \hat{r}_j)
\int_{-\infty}^{\infty}\,\frac{d\omega}{\omega}
\,\Biggl\{e^{-i\omega (t-t')} 
\left[\frac{\overrightarrow{R}_{\omega l}(r_i)\overrightarrow{R}^{*}_{\omega l}(r_j)}{1 - e^{-2\pi\omega/\kappa}} - \frac{\overleftarrow{R}_{\omega l}(r_i)\overleftarrow{R}^{*}_{\omega l}(r_j)}{e^{2\pi\omega/\kappa} - 1}\right]
\nn\\
&& - \,e^{i\omega (t-t')} 
\left[\frac{\overrightarrow{R}_{\omega l}(r_j)\overrightarrow{R}^{*}_{\omega l}(r_i)}{1 - e^{-2\pi\omega/\kappa}} - \frac{\overleftarrow{R}_{\omega l}(r_j)\overleftarrow{R}^{*}_{\omega l}(r_i)}{e^{2\pi\omega/\kappa} - 1}\right]\Biggr\}.
\label{pauli-rest-haws}
\eea
In this way one has, for the vacuum fluctuation contribution:
\bea
(\delta E)^{A}_{g,\textrm{vf}} &=& \frac{i \lambda^4}{2^{14}\pi^4}\,g_{00}(r_{A})g_{00}(r_{B})
\,\int_{-\infty}^{\infty}\,\frac{d\omega}{\omega}\int_{-\infty}^{\infty}\,\frac{d\omega'}{\omega'}
\int_{t_0}^{t}dt'\,\int_{t_0}^{t'}dt''\,\int_{t_0}^{t''}dt'''
\nn\\
&\times&\,\left[e^{-i\omega (t-t''')}H^{+}(\omega, r_A, r_B) + e^{i\omega (t-t''')}H^{+}(\omega, r_B, r_A)\right]
\nn\\
&\times&\,\left[e^{-i\omega' (t'-t'')}H^{-}(\omega', r_A, r_B) - e^{i\omega' (t'-t'')}H^{-}(\omega', r_B, r_A)\right]
\nn\\
&\times&\, \biggl[e^{i\omega_{A}(t - t')} - e^{-i\omega_{A}(t - t')}\biggr]
\biggl[e^{i\omega_{B}(t'' - t''')} - e^{-i\omega_{B}(t'' - t''')}\biggr],
\eea
where we have defined $H^{\pm}(\omega, r, r') = \overrightarrow{H}(\omega, r, r') \pm \overleftarrow{H}(\omega, r, r')$, with 
\bea
\overrightarrow{H}(\omega, r, r') &=& \sum_{l = 0}^{\infty} (2l+1)P_{l}(\hat{r} \cdot \hat{r}')\,
\overrightarrow{R}_{\omega l}(r)\overrightarrow{R}^{*}_{\omega l}(r')\left(1 +\frac{1}{e^{2\pi\omega/\kappa} - 1}\right)
\nn\\
\overleftarrow{H}(\omega, r, r') &=& \sum_{l = 0}^{\infty} (2l+1)P_{l}(\hat{r} \cdot \hat{r}')
\frac{\overleftarrow{R}_{\omega l}(r)\overleftarrow{R}^{*}_{\omega l}(r')}{e^{2\pi\omega/\kappa} - 1}.
\eea
Regarding the radiation reaction, one gets instead
\bea\label{HHrr}
(\delta E)^{A}_{g,\textrm{rr}} &=& -\frac{i \lambda^4}{2^{14}\pi^4}\,g_{00}(r_{A})g_{00}(r_{B})
\,\int_{-\infty}^{\infty}\,\frac{d\omega}{\omega}\int_{-\infty}^{\infty}\,\frac{d\omega'}{\omega'}
\int_{t_0}^{t}dt'\,\int_{t_0}^{t'}dt''\,\int_{t_0}^{t''}dt'''
\nn\\
&\times&\,\left[e^{-i\omega (t-t')}H^{-}(\omega, r_A, r_B) - e^{i\omega (t-t')}H^{-}(\omega, r_B, r_A)\right]
\nn\\
&\times&\,\left[e^{-i\omega' (t'''-t'')}H^{-}(\omega', r_A, r_B) - e^{i\omega' (t'''-t'')}H^{-}(\omega', r_B, r_A)\right]
\nn\\
&\times&\, \biggl[e^{i\omega_{A}(t - t''')} + e^{-i\omega_{A}(t - t''')}\biggr]
\biggl[e^{i\omega_{B}(t' - t'')} - e^{-i\omega_{B}(t' - t'')}\biggr].
\eea
\subsubsection{Atoms far from the black hole}
Let us evaluate the radiative energy shifts for the atom $A$ in its
ground state $|g\rangle$ in the asymptotic regions of interest,
keeping $|{\bf x}_{A} - {\bf x}_{B}|$ fixed. For $r_{A}, r_{B} \to
\infty$ and using the results discussed above, one gets
\bea\label{HHvf}
(\delta E)^{A}_{g,\textrm{vf}} &\approx& -\frac{i\lambda^4}{16}\,\int_{t_0}^{t}dt'\,\int_{t_0}^{t'}dt''\,\int_{t_0}^{t''}dt'''\,
D_{\beta}^{M}(t-t''',{\bf x}_{A}-{\bf x}_{B})\Delta^{M}(t'-t'',{\bf x}_{A}-{\bf x}_{B})\,
\nn\\
&\times&\,\sin[\omega_{0}(t - t')]\,\sin[\omega_{0}(t'' - t''')],
\eea
while $(\delta E)^{A}_{g,\textrm{rr}}$ is given by Eq.~(\ref{rr-infinity}). Note that the radiation reaction does not get any Planckian factor, since any information on the distribution function of particles is contained in the symmetric correlation function (compare with Eq.~\eqref{rr-shift}). In the above expression, $D_{\beta}^{M}$ is the thermal correlation function of the massless scalar field in Minkowski spacetime:
\bea
D_{\beta}^{M}(t-t',{\bf x}-{\bf x}') &=& \frac{1}{4\pi^2}\int_{-\infty}^{\infty} d\omega\,
\frac{\sin\left(\omega|{\bf x} - {\bf x}'|\right)}{|{\bf x} - {\bf x}'|}\,
\frac{\left(e^{-i\omega(t-t')} + e^{i\omega(t-t')}\right)}{e^{\beta\omega} - 1}
\nn\\
&=&\,\sum_{k = -\infty}^{\infty}D^{M}(t+ik\beta,{\bf x}; t', {\bf x}')
\nn\\
&=&\,\frac{1}{4\pi\beta\,|{\bf x} - {\bf x}'|}\,\left\{\coth\left[\frac{\pi[|{\bf x} - {\bf x}'| - (t-t')]}{\beta }\right] 
+ \coth\left[\frac{\pi[|{\bf x} - {\bf x}'| + (t-t')]}{\beta }\right]\right\}.
\eea
In the limit $r_{A}, r_{B} \to \infty$ one must recover the results
for the scalar Casimir-Polder energy between two static atoms at a
finite temperature $\beta^{-1} = \kappa/2\pi$, which in the present
case is just the usual Hawking temperature of the black hole. Hence, after taking the limit $t - t_0 \to \infty$ of
Eqs.~\eqref{HHrr},\eqref{HHvf},
 one finds by straightforward integration that the contributions
of vacuum fluctuations to the Casimir-Polder interaction energy are
given by 
\bea
(\delta E)^{A}_{g,\textrm{vf}} &\approx& \frac{\lambda^4}{2048\pi^4\beta\omega _0^2(\Delta r)^2}
\,\left\{4 \pi ^2 + 2 \pi  \beta  \omega _0 \left[\pi  \coth \left(\frac{\beta  \omega _0}{2}\right)
- i \Phi \left(e^{-\frac{4 \pi \Delta r}{\beta }},1,\frac{i \beta  \omega _0}{2 \pi }\right) \right. \right.
\nn\\
&+&\, \left. \left. i \Phi \left(e^{-\frac{4 \pi  \Delta r}{\beta }},1,-\frac{i \beta  \omega _0}{2 \pi }\right)\right] 
+ \beta ^2 \omega _0^2 \left[-\psi ^{(1)}\left(\frac{i \beta  \omega _0}{2 \pi }\right)
-\psi ^{(1)}\left(-\frac{i \beta  \omega _0}{2 \pi }\right) \right. \right.
\nn\\
&+&\,\left. \left. \Phi \left(e^{-\frac{4 \pi \Delta r}{\beta }},2,\frac{i \beta  \omega _0}{2 \pi }\right)
+\Phi \left(e^{-\frac{4 \pi  \Delta r}{\beta }},2,-\frac{i \beta  \omega _0}{2 \pi }\right)\right]\right\},
\eea
\end{widetext}
whereas the radiation-reaction contribution is given by
expression~(\ref{casimir-rr}). In the expression above, we have introduced
$$
\Phi(z,s,a) = \sum_{n=0}^{\infty}\frac{z^{n}}{(a+n)^s},
$$
$|z| < 1$ and $a \neq 0, -1, -2, \cdots$,  the Lerch transcendent, and
$$
\psi^{(n)}(z) = \frac{d^{n+1}}{dz^{n+1}} \ln \Gamma[z],
$$
 the polygamma function~\cite{abram} ($\Gamma(z)$ is the usual gamma function). For $a \gg 1$, one has, from the definition of the Lerch transcendent:
\bea
\Phi(z,s,a) &=& \sum_{n=0}^{\infty}\frac{z^{n}}{a^s}\left(1 - s\frac{n}{a} + \frac{s(s+1)}{2}\frac{n^2}{a^2} + \cdots \right),
\nn\\
&\approx&\,\frac{a^{-s}}{1-z} - s\left[\frac{z\,a^{-s-1}}{(z-1)^2}\right] 
\nn\\
&-&\, \frac{s(s+1)}{2}\left[\frac{z (z+1)a^{-s-2}}{(z-1)^3}\right].
\eea
In addition, one has the asymptotic formulae ($z \gg 1$)
\beq
\psi^{(1)}(z) = \frac{1}{z} + \frac{1}{2z^2} + \cdots
\eeq
and $\coth z = 1 + 2 (e^{-2z} + e^{-4z} + \cdots)$. Such results allow
us to express the Casimir-Polder energy in the limit $\beta\omega_0
\gg 1$ (low temperature): we find 
\bea
(\delta E)^{A}_{g,\textrm{vf}} &\approx& \frac{\lambda^4}{1024\pi^2\beta \omega _0^2 (\Delta r)^2}
\nn\\
&\times&\,\left[\beta  \omega _0-4 \coth \left(\frac{2 \pi  \Delta r}{\beta }\right)\right],
\label{low-T}
\eea
where we have kept only the leading-order terms in the  asymptotic expansion; in the limit $2\pi \Delta r/\beta \ll 1$, one has %
\bea
E_{\rm CP} &\approx& -\frac{\lambda^4}{512\pi^3\omega _0^2 (\Delta r)^3},
\label{low-T1}
\eea
which coincides with the leading order from~\eqref{low-T}, since the radiation-reaction contribution is negligible compared to the vacuum-fluctuation contribution. This expression  is the far-zone Casimir-Polder energy of two static atoms in Minkowski spacetime at distances where  thermal corrections are subleading since they are parametrically small in $\Delta r/\beta\ll1$. The benchmark case for this results is found again in Ref.~\cite{marino}.  On the other hand, in the limit $2\pi \Delta r/\beta \gg 1$, thermal corrections affect the scaling of the Casimir-Polder force, the reaction radiation term provides again a contribution of the same order of the vacuum fluctuation one, and therefore we find
\bea
E_{\rm CP} &\approx&
-\frac{\lambda^4}{256\pi^2\beta\omega _0^2 (\Delta r)^2},
\label{low-T2}
\eea
 which agrees once again with the thermal Casimir force computed in Minkowski space time~\cite{marino}, at the Hawking temperature $T=\kappa/2\pi$.

\subsubsection{Atoms close to the black hole}

\begin{widetext}
With $r_{A}, r_{B} \to 2GM$ and again using results from above, one has, at leading order in $r,r'\to2GM$:
\bea
(\delta E)^{A}_{g,\textrm{vf}} &\approx& -\frac{i\lambda^4}{16}\,g_{00}^2(r)
\int_{t_0}^{t}dt'\,\int_{t_0}^{t'}dt''\,\int_{t_0}^{t''}dt'''\,
\tilde{D}^{H}(t-t''',d_1,d_2)\,\tilde{\Delta}^{B}(t'-t'',d_1,d_2)
\nn\\
&\times&\,\sin[|\omega_{A}|(t - t')]\,\sin[|\omega_{A}|(t'' - t''')],
\label{vfhh-2m}
\eea
with $(\delta E)^{A}_{g,\textrm{rr}}$  given by Eq.~(\ref{rr-2m}), and
\bea
\tilde{D}^{H}(t-t',{\bf x}-{\bf x}') &=& \frac{(g_{00})^{-1/2}}{4\pi^2\Delta L\sqrt{1 + (\kappa(r) \Delta L/2)^2}}
\,\int_{-\infty}^{\infty} d\omega\,\sin\left(\omega d_{1}\right)\,\frac{\left(e^{-i\omega(t-t')} + e^{i\omega(t-t')}\right)}{e^{\beta\omega} - 1}
\nn\\
&=& \frac{(g_{00})^{-1/2}}{4\pi\beta\Delta L\sqrt{1 + (\kappa(r) \Delta L/2)^2}}
\left\{\coth\left[\frac{\pi[d_1 - (t-t')]}{\beta }\right] + \coth\left[\frac{\pi[d_1 + (t-t')]}{\beta }\right]\right\},
\label{haw-hada-2m}
\eea
\end{widetext}
\begin{figure}[t!]
\includegraphics[width=9cm]{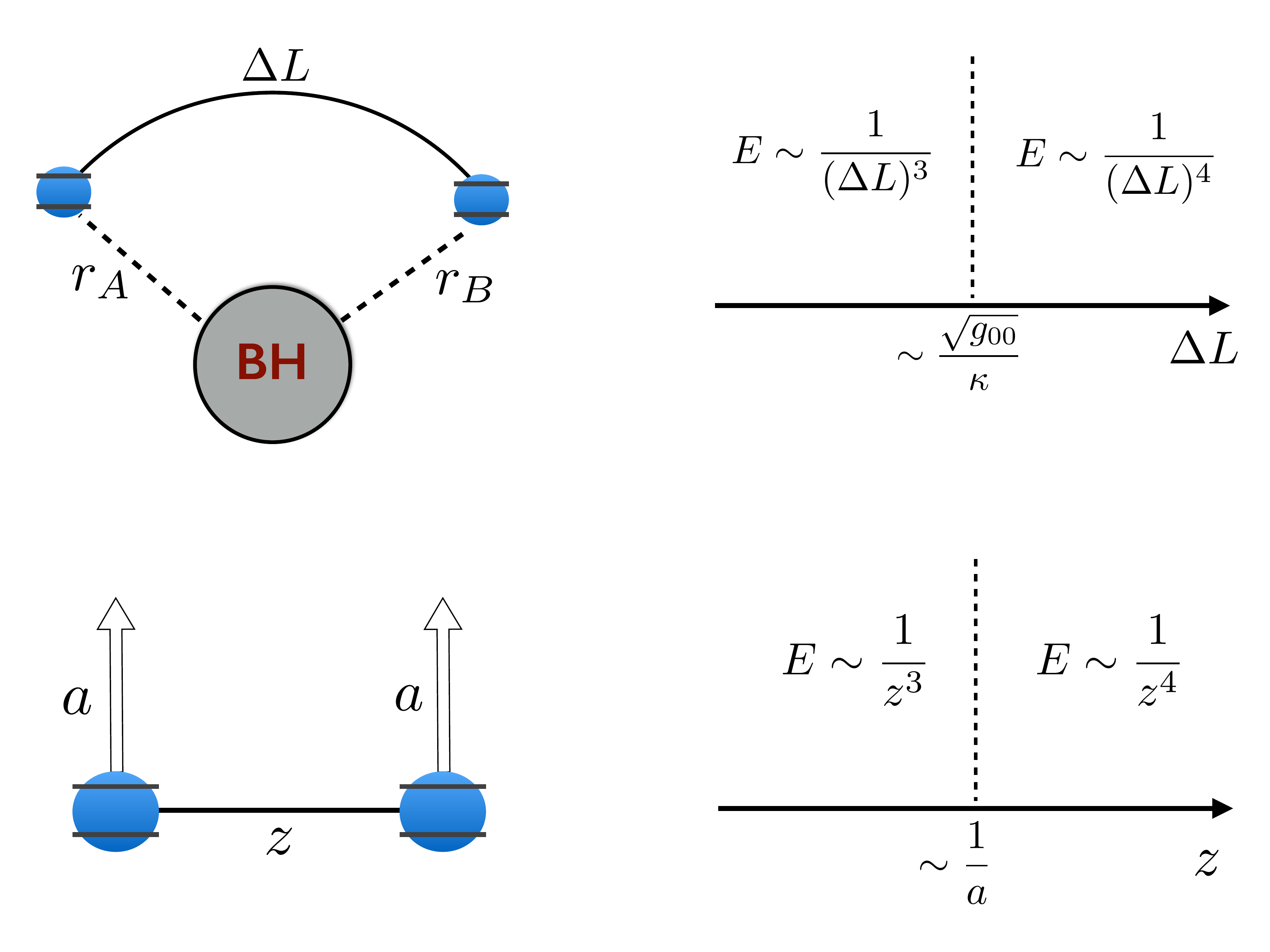}
\caption{(Color online) The  Casimir-Polder interaction between two
  atoms placed close to a black-hole horizon, displays the same
  non-thermal scaling behaviour as a function of the interatomic
  distance occurring for two relativistically uniformly accelerated
  atoms in a flat spacetime \cite{marino} (Rindler metric). The
  characteristic crossover distance is in the former case proportional
  to the inverse of the surface gravity $\kappa$, while in the second
  case to the inverse of the proper acceleration $a$. Instead of the
  thermal scaling ($1/(\Delta L)^2$ or $1/z^2$) expected from the
  thermal Hawking/Unruh radiation, the Casimir interaction decreases
  faster to zero at large distances ($1/(\Delta L)^4$ or $1/z^4$) as a
  result of further distance-dependent contributions coming from the
  non-inertial character of the background metric. For short distances
  ($\Delta L \ll \sqrt{g_{00}}/\kappa$ or $z\ll 1/a$), the Casimir
  interatomic interaction is, instead, at leading order equivalent to
  its zero-temperature expression.}  
\label{figureBH}
\end{figure}
with $\tilde{\Delta}^{B}$ given by
expression~(\ref{boul-pauli-2m}). Also in this case, there is no
signature of the thermal distribution function in the expression for
the radiation-reaction contribution. From these quantities, it is easy
to see that the radiation-reaction contribution is once again given by
expression~(\ref{casimir-rr-2m}) in this limit. On the other hand, for
the vacuum-fluctuation contribution, one has (in the limit $t - t_0
\to \infty$) 
\begin{widetext}

\bea
(\delta E)^{A}_{g,\textrm{vf}} &\approx& \frac{\lambda^4}{2048\pi^4\beta|\omega _A|^2}
\frac{g_{00}}{(\Delta L)^2\,(1 + (\kappa(r) \Delta L/2)^2)}
\,\left\{4 \pi ^2 + 2 \pi  \beta  |\omega _A| \left[\pi  \coth \left(\frac{\beta  |\omega _A|}{2}\right)
- i \Phi \left(e^{-\frac{4 \pi d_1}{\beta }},1,\frac{i \beta  |\omega _A|}{2 \pi }\right) \right. \right.
\nn\\
&+&\, \left. \left. i \Phi \left(e^{-\frac{4 \pi  d_1}{\beta }},1,-\frac{i \beta  |\omega _A|}{2 \pi }\right)\right] 
+ \beta ^2 |\omega _A|^2 \left[-\psi ^{(1)}\left(\frac{i \beta  |\omega _A|}{2 \pi }\right)
-\psi ^{(1)}\left(-\frac{i \beta  |\omega _A|}{2 \pi }\right) \right. \right.
\nn\\
&+&\,\left. \left. \Phi \left(e^{-\frac{4 \pi d_1}{\beta }},2,\frac{i \beta  |\omega _A|}{2 \pi }\right)
+\Phi \left(e^{-\frac{4 \pi d_1}{\beta }},2,-\frac{i \beta  |\omega _A|}{2 \pi }\right)\right]\right\}.
\eea
\end{widetext}
As above, in the limit $\beta\omega_A \gg 1$, the Casimir-Polder energy is given by the vacuum-fluctuation contribution,
\bea
E_{\rm CP} \approx (\delta E)^{A}_{g,\textrm{vf}} &\approx& \frac{\lambda^4}{1024\pi^2\beta|\omega _A|^2}
\frac{g_{00}}{(\Delta L)^2\,(1 + (\kappa(r) \Delta L/2)^2)}
\nn\\
&\times&\,\left[\beta  |\omega _A| - 4 \coth \left(\frac{2 \pi  d_1}{\beta }\right)\right].
\eea
For $\kappa(r) \Delta L \ll 1$, we obtain the same results as above (considering gravitational-redshift effects), namely the vacuum-fluctuation contribution to the Casimir-Polder interaction near the event horizon exhibits, at the lowest order in $\kappa(r) \Delta L$, a scaling with the interatomic distance, characteristic of the finite temperature case. However, when $\kappa(r) \Delta L \gg 1$, one gets
\beq
(\delta E)^{A}_{g,\textrm{vf}} \approx \frac{\lambda^4\,\kappa^2}{256\pi^2 |\omega _A|}\frac{1}{(\kappa(r)\Delta L)^4}
-\frac{\lambda^4 \kappa^4}{256\pi^4|\omega _A|^2}\frac{\beta}{(\kappa(r)\Delta L)^4}.
\eeq
Hence, taking into account the radiation reaction contribution given
by Eq.~(\ref{casimir-rr-2m2}), one has that 
\beq\label{non-thermal}
E_{CP} \approx -\frac{\lambda^4 \kappa^4}{256\pi^4|\omega _A|^2}\frac{\beta}{(\kappa(r)\Delta L)^4}.
\eeq

The scaling of the Casimir energy~\eqref{non-thermal}  is in line with the conclusions of Ref.~\cite{marino} where the scaling of the Casimir-Polder force has been computed for two uniformly relativistic accelerating atoms. 
 In particular, for interatomic distances larger than the typical length scale
 $1/\kappa(r)$, where the metric (and accordingly the field
 correlation functions) displays a strong noninertial character, the
 thermal scaling is deformed, and a novel scaling form for  the
 Casimir-Polder interaction energy sets in. In close parallel, in
 Ref.~\cite{marino} it has been shown that this non-thermal scaling,
 $1/z^4$, of the Casimir interaction as a function of the interatomic
 distance $z$, occurs at distances larger than $\sim1/a$, with $a$
  the proper acceleration of the two atoms. 
The feature of having analogue scalings and the acceleration replaced by the surface gravity  is expected on the basis of the equivalence of the Rindler metric with the Schwarzschild one, when the atoms are located close to the black hole horizon.
The interpretation follows again the features discussed in the case of
the Boulware vacuum: On the top of a thermal scaling regulated by the
Hawking temperature $T$, additional factors depending on interatomic
distance enter the Casimir interaction to implement the
noninertial character of the background metric. Once again, this is a
direct consequence of a change in the scaling of response ($\Delta$)
and correlation functions ($D$), which is provoked by curvature
corrections to the correlation functions at large enough distances
where gravity effects are more pronounced. 

The analogy between the nonthermal character of Casimir interaction
in a Rindler and Schwarzschild background  constitutes the central
result of our work, and we summarize a comparison between these two cases in
Fig.~\ref{figureBH}.

\subsection{Unruh vacuum}
\label{Unruh}

The Unruh vacuum state is the adequate choice of vacuum state relevant for the gravitational collapse of a massive body~\cite{sciama,unruh}. At spatial infinity, this vacuum depicts an outgoing flux of black-body radiation at the black-hole temperature.  

By inserting Eq.~(\ref{unruh}) from the Appendix in Eqs.~(\ref{hada}) and~(\ref{pauli}) one obtains the associated symmetric and response correlation functions, respectively. Not surprisingly, we obtain similar results as above within the same asymptotic limits.
For instance, for $r_{A}, r_{B} \to \infty$, one can easily
prove that $(\delta E)^{A}_{g,\textrm{vf}}$ is given by
Eq.~(\ref{vf-infinity}), whereas $(\delta E)^{A}_{g,\textrm{rr}}$ is
given by expression~(\ref{rr-infinity}). In other words, one obtains
the same results as in the Boulware vacuum at spatial infinity. In
turn, with $r_{A}, r_{B} \to 2GM$, one has that $(\delta
E)^{A}_{g,\textrm{vf}}$ and $(\delta E)^{A}_{g,\textrm{rr}}$ are given,
respectively, by Eqs.~(\ref{vfhh-2m}) and~(\ref{rr-2m}). That is, one
obtains the same results as in the Hartle-Hawking vacuum near the
event horizon.

\section{Conclusions and Perspectives}
\label{conclude}

We have discussed the contributions of vacuum fluctuations and
radiation reaction to the Casimir-Polder forces between two identical
atoms in Schwarzschild spacetime.
We have shown how the distance-dependent radiative shifts of
atoms in their ground states are modified when the atoms are placed near and far
away from the black hole as well as when the quantum field is prepared
in the Boulware, Hartle-Hawking, and Unruh vacuum states.  
Our findings generalize, in particular, the  mechanism discussed in
Ref.~\cite{marino} for two uniformly relativistic accelerated atoms:
The Casimir-Polder interaction exhibits a transition between different
scaling behaviors (thermal and nonthermal like) at a characteristic
length associated with the mass of the black hole. This effect is
pronounced close  to the event horizon, and it originates from the
noninertial character of the background metric, which provides further
distance-dependent corrections to the otherwise expected thermal (at
Hawking temperature) scaling of the Casimir interaction energy. 
Furthermore, close to the black hole, where the Schwarzschild metric
takes the form of the Rindler line element, we find the same
qualitative scaling as was found in Ref.~\cite{marino} for two
relativistically uniformly accelerated two-level atoms.  

There have been several investigations of quantum electrodynamic effects in a curved spacetime. Indeed, there is a number of discussions of the behavior of a scalar field (such as the Higgs particle) in the vicinity of strong gravitational sources~\cite{onofrio1}. 
In turn, Ref.~\cite{referee} considers the Higgs self-interaction in a perturbed FRW metric. On the other hand, proposals highlighting the potential of spectroscopic measurements near the surface of white dwarfs and neutron stars can be found in Refs.~\cite{onofrio2}. In a similar spirit, radiative shifts of matter surrounding a black hole might be significantly altered by the qualitative distance-dependent corrections discussed in this work. In this way the present results provide an indirect confirmation of corrections to the scaling of Casimir-Polder forces in accelerated backgrounds.

The formulae for the vacuum fluctuation and radiation reaction terms
at fourth order in perturbation theory constitute a promising tool to compute Casimir interactions
for ground-state atoms in other more complicated settings. 

It would be interesting to generalize our results to
other situations where quantum aspects of the
gravitational field are of relevance.
The first example is the study of gedanken experiments like the one
discussed in \cite{UW82}, in which a box is lowered towards the event
horizon. Quantum effects are there important to guarantee the validity
of the Generalized Second Law of black hole mechanics.   
The second example is the Kerr black
hole (see e.g. \cite{wheeler}). In contrast to a Schwarzschild black
hole, it has a region called ergosphere in which static observers
cannot exist. The calculation of Casimir-Polder energies near or in
this region are of  interest, but could also be of astrophysical
relevance because observed black holes (such as the supermassive
black hole with $4.3 \times 10^6 M_{\odot}$ in the center of the
Milky Way) all have accretion disks of matter around them. 
Finally, the behavior of Casimir-Polder forces near cosmological
horizons (de~Sitter case) or in situations with both cosmological and black
hole horizons (Schwarzschild-de~Sitter case) \cite{GH77,FGK} 
could turn out to be of conceptual interest. All of this would boost
out knowledge about the intriguing features that appear when quantum
theory and gravitational physics are intertwined.

\section*{acknowlegements}

CK and JM thank Bill Unruh for an interesting discussion.
JM thanks A. Noto, R. Passante, L. Rizzuto, S. Spagnolo for collaboration and discussions on closely related research topics.
This work was partially supported by `Conselho Nacional de
Desenvolvimento Cientifico e Tecnol{\'o}gico' (CNPq, Brazil). 
JM acknowledges support from the Alexander von Humboldt foundation.

\appendix

\section{Correlation functions of the scalar field in Schwarzschild
  spacetime} 
\label{A}

Here we present the correlation function of the quantum scalar field
in Schwarzschild spacetime (for more details we refer the reader to
\cite{frolov,candelas} and references cited therein). The
Lagrangian density is given by  
\beq
S = \frac{1}{2}\,\int d^4 x\,\sqrt{-g}\,g^{\alpha\beta}\varphi_{,\alpha}\varphi_{,\beta}.
\label{sc-field}
\eeq
In the exterior region of Schwarzschild spacetime, a complete set of normalized basis functions for the massless scalar field is
\beq
u_ {\omega l m n}(x) =\frac{1}{\sqrt{4\pi\omega}}\, R^{(n)}_{\omega l}(r) Y_{lm}\,e^{-i\omega t},
\eeq
where the label $n$ distinguishes between modes incoming from past null infinity ${\cal J}^{-}$ (hereafter denoted by $n = \leftarrow$) and modes going out from the past horizon ${\cal H}^{-}$ (hereafter denoted by $n = \rightarrow$). One has the asymptotic forms:
\bea
\overrightarrow{R}_{\omega l}(r) &\sim& r^{-1}\,e^{i\omega r_{*}} + 
\overrightarrow{{\cal R}}_{l}(\omega)\,r^{-1}\,e^{-i\omega r_{*}},\,\,\,r \to 2GM
\nn\\
\overrightarrow{R}_{\omega l}(r) &\sim& \overrightarrow{{\cal T}}_{l}(\omega)\,r^{-1}\,e^{i\omega r_{*}},\,\,\,r \to \infty
\nn\\
\overleftarrow{R}_{\omega l}(r) &\sim& \overleftarrow{{\cal T}}_{l}(\omega)\,r^{-1}\,e^{-i\omega r_{*}},\,\,\,r \to 2GM
\nn\\
\overleftarrow{R}_{\omega l}(r) &\sim& r^{-1}\,e^{-i\omega r_{*}} + 
\overleftarrow{{\cal R}}_{l}(\omega)\,r^{-1}\,e^{i\omega r_{*}},\,\,\,r \to \infty,
\label{asymp-scalar}
\eea
where $r_{*} = r + 2GM\ln(r/2GM - 1)$ is the Regge-Wheeler tortoise coordinate, and ${\cal R}$ and ${\cal T}$ are the usual reflection and transmission coefficients, respectively, with the following properties
\bea
\overrightarrow{{\cal T}}_{l}(\omega) &=& \overleftarrow{{\cal T}}_{l}(\omega) = {\cal T}_{l}(\omega),
\nn\\
|\overrightarrow{{\cal R}}_{l}(\omega)| &=& |\overleftarrow{{\cal R}}_{l}(\omega)|,
\nn\\
1 - |\overrightarrow{{\cal R}}_{l}(\omega)|^2 &=& 1 - |\overleftarrow{{\cal R}}_{l}(\omega)|^2 = |{\cal T}_{l}(\omega)|^2,
\nn\\
\overrightarrow{{\cal R}}_{l}^{*}(\omega){\cal T}_{l}(\omega) &=& - {\cal T}^{*}_{l}(\omega)\overleftarrow{{\cal R}}_{l}(\omega).
\eea
The positive frequency Wightman functions associated with the Boulware
vacuum $|0_B\rangle$, the Hartle-Hawking vacuum $|0_H\rangle$, and the
Unruh vacuum $|0_U\rangle$ are given, respectively, by 
\begin{widetext}
\bea
\langle 0_B|\varphi(x)\varphi(x')|0_B\rangle &=&\sum_{l m}\,\int_{0}^{\infty}\,\frac{d\omega}{4\pi\omega}\,
e^{-i\omega (t-t')} Y_{lm}(\theta,\phi)Y^{*}_{lm}(\theta',\phi')
\nn\\
&\times&\,\left[\overrightarrow{R}_{\omega l}(r)\overrightarrow{R}^{*}_{\omega l}(r') + 
\overleftarrow{R}_{\omega l}(r)\overleftarrow{R}^{*}_{\omega l}(r')\right],
\eea
\bea
\langle 0_H|\varphi(x)\varphi(x')|0_H\rangle &=& \sum_{l m}\,\int_{-\infty}^{\infty}\,\frac{d\omega}{4\pi\omega}\,
\Biggl[e^{-i\omega (t-t')} Y_{lm}(\theta,\phi)Y^{*}_{lm}(\theta',\phi')
\frac{\overrightarrow{R}_{\omega l}(r)\overrightarrow{R}^{*}_{\omega l}(r')}{1 - e^{-2\pi\omega/\kappa}}
\nn\\
&+& e^{i\omega (t-t')} Y^{*}_{lm}(\theta,\phi)Y_{lm}(\theta',\phi')
\frac{\overleftarrow{R}^{*}_{\omega l}(r)\overleftarrow{R}_{\omega l}(r')}{e^{2\pi\omega/\kappa} - 1}\Biggr],
\eea
and
\bea
\langle 0_U|\varphi(x)\varphi(x')|0_U\rangle &=& \sum_{l m}\,\int_{-\infty}^{\infty}\,\frac{d\omega}{4\pi\omega}\,
e^{-i\omega (t-t')} Y_{lm}(\theta,\phi)Y^{*}_{lm}(\theta',\phi')
\nn\\
&\times&\,\left[\frac{\overrightarrow{R}_{\omega l}(r)\overrightarrow{R}^{*}_{\omega l}(r')}{1 - e^{-2\pi\omega/\kappa}} + \theta(\omega)\overleftarrow{R}_{\omega l}(r)\overleftarrow{R}^{*}_{\omega l}(r')\right],
\label{unruh}
\eea
where $\kappa = 1/4GM$ is the surface gravity of the black
hole~\cite{birrel}.

 Let us now present the mode summations in the asymptotic regions $r
 \to 2GM$ and $r \to \infty$. At fixed radial distances $r$ and $r'$,
 the correlation function of the field in the Boulware vacuum can be
 written as
\bea
\langle 0_B|\varphi(x)\varphi(x')|0_B\rangle &=& \frac{1}{16\pi^2}\sum_{l=0}^{\infty}\,\int_{0}^{\infty}\,
\frac{d\omega}{\omega}\,e^{-i\omega (t-t')} (2l + 1)\,P_{l}(\hat{r} \cdot \hat{r}')
\nn\\
&\times&\,\left[\overrightarrow{R}^{(1)}_{\omega l}(r)\overrightarrow{R}^{(1 *)}_{\omega l}(r') + \overleftarrow{R}^{(1)}_{\omega l}(r)\overleftarrow{R}^{(1 *)}_{\omega l}(r')\right].
\label{b-sph-scalar}
\eea
where we have used the addition theorem for the spherical harmonics, and $\hat{r}$ and $\hat{r}'$ are two unit vectors with spherical coordinates $(\theta, \phi)$ and $(\theta', \phi')$. In general, one has 
\beq
\sum_{l=0}^{\infty}(2l + 1)\,P_{l}(\hat{r} \cdot \hat{r}')\overrightarrow{R}_{\omega l}(r)\overrightarrow{R}^{*}_{\omega l}(r') \approx \sum_{l=0}^{\infty}\frac{(2l + 1)\,P_{l}(\hat{r} \cdot \hat{r}')|{\cal T}_{l}(\omega)|^2 e^{i\omega (r_{*} - r'_{*})}}{r r'},\,\,\, r,r' \to \infty.
\label{asymp-inf-scalar}
\eeq
\end{widetext}
and for ${\bf x} = {\bf x}'$ we get ($P_{l}(1) = 1$):
\beq
\sum_{l=0}^{\infty} (2l + 1)|\overrightarrow{R}_{\omega l}|^2 \approx \sum_{l=0}^{\infty}\frac{(2l+1)|{\cal T}_{l}(\omega)|^2}{r^2},\,\,\, r \to \infty.
\label{asymp-inf2-scalar}
\eeq
In order to estimate the remaining sum, it is an important benchmark to recall that  the above correlation function  should agree at large radii with the correlation function of the scalar field in the Minkowski vacuum. Therefore, for $r, r' \to \infty$, one gets
\bea
&&\sum_{l=0}^{\infty}(2l + 1)\,P_{l}(\hat{r} \cdot \hat{r}')\overleftarrow{R}_{\omega l}(r)\overleftarrow{R}^{*}_{\omega l}(r')
\nn\\
&\approx& 4\omega\,\frac{\sin\left(\omega|{\bf x} - {\bf x}'|\right)}{|{\bf x} - {\bf x}'|},\,\,\, r, r' \to \infty,
\label{asymp-inf3-scalar}
\eea
and in conclusion, for ${\bf x} = {\bf x}'$
\beq
\sum_{l=0}^{\infty} (2l + 1)|\overleftarrow{R}_{\omega l}|^2 \approx 4\omega^2,\,\,\, r \to \infty.
\label{asymp-inf4-scalar}
\eeq
Let us evaluate the mode sums in the region $r \approx 2GM$. We begin by defining $\zeta^2 = r/2GM - 1$ and $q = 4 GM \omega$. With these definitions, one can prove that $\overrightarrow{R}_{\omega l}$ obeys an equation that has the following approximate form:
\beq
\left[\zeta^2\frac{d^2}{d\zeta^2} + \zeta\frac{d}{d\zeta} + \left(q^2 - (2l\zeta)^2\right)\right]\overrightarrow{R}_{\omega l}(\zeta) = 0,
\eeq
where we have approximated $l(l+1)\zeta^2 \approx (l\zeta)^2$ since $\zeta \sim 0$. This is just the usual Bessel differential equation whose general solution can be expressed in terms of the modified Bessel functions:
\beq
\overrightarrow{R}_{\omega l}\big|_{r \to 2GM} \sim c_{l} K_{iq}(2l\zeta) + d_{l} I_{-iq}(2l\zeta) . 
\label{sol-varphi-scalar}
\eeq
\pagebreak
\begin{widetext}
As $l \to \infty$ for fixed $\zeta$, the radial function tends to zero; $r$ lies in the region in which the effective potential for the radial function is large. Hence $d_{l}$ is an exponentially small function of $l$ for large $l$ and the second term in equation~(\ref{sol-varphi-scalar}) may be neglected in comparison with that of the first term in~(\ref{sol-varphi-scalar}). The coefficient $c_{l}$ may be determined by comparing the asymptotic result $z \to 0$ for $K_{\nu}(z)$ with the asymptotic solution
$$
\overrightarrow{R}_{\omega l}(r) \sim e^{i\omega r_{*}}\,r^{-1} + \overrightarrow{{\cal R}}_{l}(\omega)e^{-i\omega r_{*}}\,r^{-1},\,\,\,r \to 2GM.
$$
One finds that
\beq
c_{l} \sim \frac{e^{iq/2}\,l^{-iq}}{GM\Gamma(-iq)}.
\eeq
Therefore, at leading order we have
\bea
&&\sum_{l=0}^{\infty}(2l + 1)\,P_{l}(\hat{r} \cdot \hat{r}')\overrightarrow{R}_{\omega l}(r)
\overrightarrow{R}^{*}_{\omega l}(r') \approx 
\frac{1}{GM^2\Gamma(iq)\Gamma(-iq)}\sum_{l}^{\infty}(2l + 1)P_{l}(\cos\gamma)\,K_{iq}(2l\zeta)K_{iq}(2l\zeta')
\nn\\
&&\approx \frac{8GM\omega\sinh(4\pi GM \omega)}{\pi GM^2}\int_{0}^{\infty}dl\,l\,J_{0}(l\gamma)\,K_{iq}\left(2l\sqrt{g_{00}(r)}\right)K_{iq}\left(2l\sqrt{g_{00}(r')}\right),\,\,\, r, r' \to 2GM
\eea
where $g_{00} = (1 - 2GM/r)$, $\cos\gamma = \hat{r} \cdot \hat{r}' =
\cos\theta\cos\theta' + \sin\theta\sin\theta'\cos(\phi - \phi')$, and
we have used~\cite{abram} 
$$
\Gamma(iq)\Gamma(-iq) = \frac{\pi}{q\sinh(\pi q)},
$$
together with the asymptotic result:
$$
P_{\nu}\left(\cos\frac{x}{\nu}\right) \approx J_{0}(x) + {\cal O}(\nu^{-1}),
$$
in which $J_{\mu}(x)$ is a Bessel function of the first kind. Considering that $r \approx r'$ (but ${\bf \hat{r}} \neq {\bf \hat{r}}'$), one may resort to the result~\cite{prud}:
$$
\int_{0}^{\infty}dx\,x J_0(a x) [K_{i q}(2 b x)]^2 = \frac{i \pi  \text{csch}(\pi  q) \sin \left(2 q \sinh^{-1}\left(\frac{a}{4b}\right)\right)}{a^2 \sqrt{-\frac{16 b^2}{a^2}-1}},
$$
where we assume a small positive imaginary part for $a$ so that the integral converges. Therefore, as a next step we find
\bea
&&\sum_{l=0}^{\infty}(2l + 1)\,P_{l}({\bf \hat{r}} \cdot {\bf \hat{r}}')\overrightarrow{R}_{\omega l}(r)\overrightarrow{R}^{*}_{\omega l}(r') \approx \frac{4\omega }{\sqrt{g_{00}}}\,\frac{\sin\left[(2\omega/\kappa) \sinh^{-1}\left(\frac{\kappa(r) \Delta L}{2}\right)\right]}{\Delta L\sqrt{1 + (\kappa(r) \Delta L/2)^2}},\,\,\, r, r' \to 2GM,
\nn\\
\label{asymp-2m-scalar2}
\eea
where $\kappa(r) = \kappa/\sqrt{g_{00}(r)}$ and $\Delta L = r_s\gamma$
are defined as above. For ${\bf x} = {\bf x}'$ we find
\beq
\sum_{l=0}^{\infty}(2l + 1)\,|\overrightarrow{R}^{(1)}_{\omega l}|^2 \approx \frac{4\omega^2}{g_{00}},\,\,\, r \to 2GM,
\label{asymp-2m2-scalar}
\eeq
where we have used that~\cite{prud}
$$
\frac{2}{\Gamma(iq)\Gamma(-iq)}\int_{0}^{\infty}dt\,t\,[K_{iq}\left(2tx\right)]^2 = \frac{q^2}{4 x^2}.
$$
The other mode sum in the region $r \to 2GM$ can be easily estimated:
\beq
\sum_{l=0}^{\infty}(2l + 1)\,P_{l}(\hat{r} \cdot \hat{r}')\overleftarrow{R}_{\omega l}(r)\overleftarrow{R}^{*}_{\omega l}(r') \approx \sum_{l=0}^{\infty}\frac{(2l+1)P_{l}(\hat{r} \cdot \hat{r}')|{\cal T}_{l}(\omega)|^2 e^{-i\omega (r_{*} - r'_{*})}}{(2GM)^2},\,\,\, r, r' \to 2GM,
\label{asymp-2m3-scalar}
\eeq
which for ${\bf x} = {\bf x}'$ reads
\beq
\sum_{l=0}^{\infty} (2l + 1)|\overleftarrow{R}_{\omega l}|^2 \approx \sum_{l=0}^{\infty}\frac{(2l+1)|{\cal T}_{l}(\omega)|^2}{(2GM)^2},\,\,\, r \to 2GM.
\label{asymp-2m4-scalar}
\eeq

\end{widetext}

\end{document}